\documentclass{article}
\usepackage[utf8]{inputenc}
\usepackage[normalem]{ulem}
\usepackage{amsmath,amssymb,amsfonts}
\usepackage{mathtools}
\usepackage{physics}
\usepackage{relsize}
\usepackage{multirow}
\usepackage[font=small,labelfont=bf]{caption}
\usepackage{subcaption}
\usepackage{makecell}
\usepackage{booktabs}
\usepackage{cite}
\usepackage{graphicx}
\usepackage[export]{adjustbox}
\usepackage{textcomp}
\usepackage{xcolor}
                     
\usepackage{makecell}
\usepackage{enumitem}
\usepackage{lineno}
\usepackage{float}
\usepackage[compact]{titlesec}

\DeclareUnicodeCharacter{2212}{-}
\title{Deep Learning assisted microwave-plasma interaction based technique for plasma density estimation}
\date{ }
\author{P. Ghosh$^1$, B. Chaudhury$^1$, S. Purohit$^2$, V. Joshi$^1$, A. Kothari$^1$, D. Shetranjiwala$^1$
\thanks{Corresponding Author :pratik\_ghosh@daiict.ac.in; bhaskar\_chaudhury@daiict.ac.in}\\
\emph{Group in Computational Science and HPC} \\ \emph{$^1$DA-IICT, Gandhinagar - 382007 , India.}\\
\emph{$^2$Institute for Plasma Research, Gandhinagar, India, 382428}}

\begin{document}



\maketitle
\begin{abstract}
The electron density is a key parameter to characterize any plasma. Most of the plasma applications and research in the area of low-temperature plasmas (LTPs) are based on the accurate estimations of plasma density and plasma temperature. The conventional methods for electron density measurements offer axial and radial profiles for any given linear LTP device. These methods have major disadvantages of operational range (not very wide), cumbersome instrumentation, and complicated data analysis procedures. The article proposes a Deep Learning (DL) assisted microwave-plasma interaction-based non-invasive strategy, which can be used as a new alternative approach to address some of the challenges associated with existing plasma density measurement techniques. The electric field pattern due to microwave scattering from plasma is utilized to estimate the density profile. The proof of concept is tested for a simulated training data set comprising a low-temperature, unmagnetized, collisional plasma. Different types of symmetric (Gaussian-shaped) and asymmetrical density profiles, in the range $10^{16}-10^{19}$ m$^{-3}$, addressing a range of experimental configurations have been considered in our study. Real-life experimental issues such as the presence of noise and the amount of measured data (dense vs sparse) have been taken into consideration while preparing the synthetic training data-sets.  The DL-based technique has the capability to determine the electron density profile within the plasma. The performance of the proposed deep learning-based approach has been evaluated using three metrics- structural similarity index (SSIM), root mean square logarithmic error (RMSLE), and mean absolute percentage error (MAPE). The obtained results show promising performance in estimating the 2D radial profile of the density for the given linear plasma device and  affirms the potential of the proposed ML-based approach in plasma diagnostics.
\end{abstract}

\vspace{2pc}
\noindent{\it Keywords}: Plasma diagnostics, microwave plasma interaction, plasma density estimation, deep learning, low temperature plasmas.
\section{Introduction}
The accurate diagnostics of low-temperature plasmas \cite{ltpdiagnostics3} is of eminence significance due to its wide array of applications, especially in microelectronics, medicine, surface engineering, packaging, biomedical research, material synthesis, pollutant degradation, chemical conversion, propulsions systems, electronic switching devices, automotive, luminous systems, and many more \cite{ltpgeneralbook,ltpgeneral1,ltpgeneral2,plasmaroadmap2012,plasmaroadmap2017,plasmaroadmap2022,mesbah,ltpgeneral3}. Low temperature plasmas can be created at various pressures, with typical ionization degrees of $10^{−6}–10^{−4}$, characteristic electron energies of a few eV to 10 eV and electron density typically from $10^{14}–10^{23}$ m$^{-3}$\cite{plasmaroadmap2017,plasmaroadmap2012,plasmaroadmap2022}. Any plasma is characterized by several physical parameters; however, the electron density ($n_{e}$) and electron temperature ($T_{e}$) are the two fundamental parameters often employed, as the plasma properties are majorly influenced by the electron dynamics \cite{ref15,ref16,vishrut}. These two parameters are important as they directly impact the plasma stability and physical or chemical properties. Plasma temperature gives a realization of the average energy carried by the plasma, whereas the electron density gives information about the available number of particles having such average energy within the given plasma volume \cite{ref15,ref16,ref17}. Accurate diagnosis and precise understanding of the spatial distribution of ($n_{e}$) and ($T_{e}$) is a prerequisite for any experimental investigation or application development \cite{ref18}. The spatial distribution or profile can change temporally.
Plasma electron density determination is carried out by some of the well-established approaches. Comprehensive reviews on different plasma diagnostics techniques can be found in the existing literature \cite{ltpdiagnostics2,ltpdiagnostics3,ltpdiagnostics4}. 

Langmuir probe, an invasive technique, is widely used for the investigation of electron characteristics in plasma; however, the mathematical theory by which the density is obtained from the Langmuir probe data (current density vs. applied potential) is quite cumbersome. The lack of a general analytic theory for any arbitrary values of density, difficulty in interpretation due to the presence of RF fields, and other issues limit the application of the Langmuir probe in plasma diagnostics. Stark broadening measurements for low $'Z’$ ionized species serve for plasma electron density ($n_e$) estimation, which typically exceeds $10^{14}$ cm$^{-3}$. The $H_{\beta}$ line emission is majorly employed for $n_e$ estimation as it has significant stark broadening for the high-pressure plasma \cite{ref19,ref20}. $H_{\alpha}$ and $H_{\gamma}$ are also employed for density estimation; however, plagued with higher self-absorption cross-sections for high-pressure plasma \cite{ref21}. These measurements are line integrated and require measurements from different locations and tomographic reconstruction for $n_e$ profile estimation \cite{ref22}. Microwave interferometry, as well as the CO2-laser heterodyne interferometry, are extensively employed in diagnostics for $n_{e}$ estimation \cite{ref23,ref24}. This method has an advantage over the line-broadening method; that is, this method applies to densities lower than $10^{13}$ cm$^{-3}$. It is noteworthy that the phase shift detection is influenced by the thermal effect, and the measurements are line integrated; therefore, constraints the $n_{e}$ profile measurement. Microwave reflectometry is also an interesting method for the estimation of $n_{e}$ \cite{ref27}, where the experimental data are fitted to the results of a numerical calculation code derived from a refined electromagnetic model. 
Microwave reflectometry uses the principle of reflection of electromagnetic waves from the target, such as gaseous plasma. Previously used in ionospheric plasma study   \cite{EMazzucato1998}, it uses group delay of the reflected microwave to correlate it with the plasma density determination. Later, the technique has been used extensively in the diagnosis of Tokamak plasmas \cite{EMazzucato1998}. The method requires sophisticated data processing and fitting. 
The Thomson scattering diagnostic (TSD) is an advanced plasma $n_{e}$ and $T_{e}$ profile determination diagnostic \cite{ref25,ref26}, as TSD offers local measurements, no line-integrated measurements. TSD works on the principle of elastic scattering of the light on the free plasma electrons, and this is measured as the Doppler broadening due to the electron velocity with the assumption of electron distribution of Maxwell-Boltzmann distribution. The major limitation of TSD performance is weak scattering signals along with the presence of stray light. However, proper filtering enables the capture of weak scattering signals. The time resolution is also an issue for small-lived plasma discharges, as the highest possible time resolution reported is of the order of a few microseconds. The requirement of real-time measurements of the plasma electron density as well as the profile information, is critical to high-pressure plasma applications.

The mentioned diagnostic systems have their inherent limitations as well as operational issues. Moreover, accurate data processing is required for the realization of the spatial and temporal $n_{e}$ profile. Space constraints associated with plasma experiments, efficient data processing requirements, and limited operational range are some of the reasons which advocate for a new non-invasive methodology, which can address mentioned shortfalls of the current approaches. The ML/DL-based techniques have started finding applications in both fusion and low-temperature plasma experiments for designing different diagnostics\cite{mesbah,gidon,Bonzanini,vishrut,Samuell,wang,niharika}. The ML and deep neural network (DNN) based techniques have already been explored in EM wave propagation and other EM interaction studies \cite{AndreaM2019}, mainly due to the effectiveness and potentialities of such DNNs as a powerful computational tool with very high computational efficiency. 
Recent applications of convolutional neural networks (CNN) based DL models have been proposed \cite{Mihir2022etal}, to predict the scattered microwave E-field from the low temperature, collisional and un-magnetized plasmas. Physics-informed Neural networks (PINN) have started gaining interest due to applications in inverse EM problems \cite{MaxwellNet2022}. To this end, the DL-assisted techniques based on EM-plasma interactions has the potential to be an important area to be explored in developing new plasma diagnostic technique by utilizing the existing literature from both areas. As a proof of concept, this article proposes a DL-assisted microwave reflectometry-based strategy that has the capability to determine the electron density in a low-temperature plasma along with its complete profile. 

A brief overview of microwave-plasma interaction, based on which traditional microwave-based plasma diagnostic techniques are designed, is presented in section II, which is followed by a detailed account of the proposed machine learning (ML) assisted microwave-plasma interaction-based approach for estimation of plasma electron density is given in section III. The section touches upon different aspects of synthetic data generation for testing via 2D electromagnetic fluid simulation. Section IV contains the employed deep learning architecture and machine learning aspects. The major finding of this work is presented in the results and discussion, section V. A summary has been drawn in section VI.

\section{Microwave-plasma interaction based technique for plasma diagnostics}

The plasma profile parameter determination using microwave-based diagnostics generally uses either microwave transmission or microwave reflection-assisted techniques \cite{MYang2021MWreflectpaper}; both techniques use the principle of EM wave-plasma interaction.
The plasma-wave interaction is one of the interesting research areas mainly to study the wave propagation characteristics within and outside the plasma. 
When an EM wave such as a microwave is launched in a weakly ionized unmagnetized plasma, it is subjected to scattering as well as absorption. Loss of EM wave energy due to energy transfer to the charged particles in plasma and subsequently to neutral particles by elastic/ inelastic collisions leads to absorption. Wave scattering is determined by the density variations within the plasma.
The plasma-wave interaction, as discussed, depends on the complex dielectric permittivity of plasma. For a collisional plasma, the complex relative dielectric permittivity can be expressed as :
\begin{eqnarray} \label{eq1}
    \epsilon (\omega)=\left(1-\frac{\omega_{p}^{2}}{\omega^2+\nu_{m}^{2}}\right)-i\left(\frac{\omega_{p}^{2}}{\omega^2+\nu_{m}^{2}}\right)\left( \frac{\nu_{m}}{\omega}\right)
\end{eqnarray}
The real part in the above equation decides the permittivity, which is denoted by $\epsilon_{r}=1-\left( \omega_{p}^{2}/(\omega^2+\nu_{m}^{2})\right)$ and the conductivity is determined by the imaginary part $\epsilon_i= \big(\nu_m/\omega\big)\bigg( \omega_{p}^{2}/(\omega^{2}+\nu_{m}^{2})\bigg)$. The conductivity is given by the formula, $\sigma=\epsilon_{0} \nu_{m}\left( \omega_{p}^{2}/(\omega^2+\nu_{m}^{2})\right)$, where $\omega_p=(n_{e} e^{2}/m_e \epsilon_0)^{1/2}$ is the plasma frequency, $\omega$ is the wave angular frequency, $\nu_{m}$ is the
electron-neutral collision frequency, $n_e$ is the local electron density that varies with position, $e$ and $m_e$ represents the electron charge and mass respectively. 
\begin{figure}[!htbp]
\centering
\vspace{-3mm}
\includegraphics[width=\textwidth]{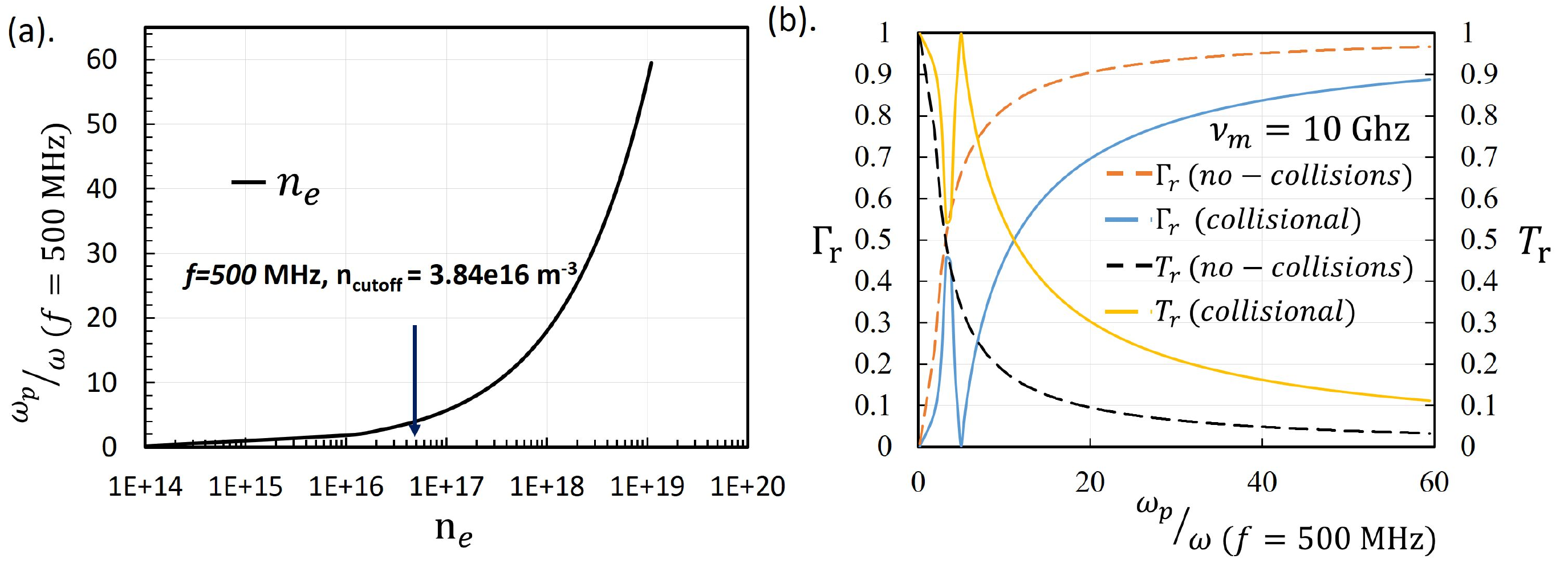}~
\caption{
The plot of the variation  (a).
$\omega _p/\omega$ vs. $n_e$ for fixed $\omega$(or \textit{f}=500 MHz )
correspond to a microwave of f= 500 MHz, and an increasing plasma density profile $1e14-1e19$ m$^{-3}$, (b).
the real component of the reflection coefficient( $\Gamma_r$) and transmission coefficient ( $T_r$) for collisional and non-collisional unmagnetized plasma is shown.}
\label{Fig:Eprandref}
\end{figure}
The complex permittivity results in plasma behaving as a debye dispersive media, which response to the incident EM wave with varying dielectric properties based on the wave frequency and local density. The dispersive nature of plasma can be explained through the propagation vector ($k=(\omega/c) n$), where c is the speed of light, n is the complex refractive index, $n=n_{r}+jn_{i}$. $n$ is related by $n=\sqrt{\epsilon(\omega)}$. The propagation constant in the plasma can be expressed by,
$k=k_{r}+jk_{i}$. Where the $Re\{k\}= k_{r}$, decides the wave propagation through the medium, also termed as phase shift constant, and $Im\{k\}=k_{i}$ decides the attenuation of the wave as it propagates through the plasma medium also termed as the attenuation coefficient. The $k_{r}$ and $k_{i}$ can be expressed in terms of the real and imaginary permittivity as, 
\begin{eqnarray} \label{eq1}
k_{r}=\bigg(\frac{\omega}{c}\bigg)\sqrt{\frac{1}{2}(\epsilon_r)+\frac{1}{2}\sqrt{\big( \epsilon_r\big)^{2}+\big( \epsilon_i\big)^{2}}}
\end{eqnarray}
\begin{eqnarray}
    k_{i}=\bigg(\frac{\omega}{c}\bigg)\sqrt{-\frac{1}{2}(\epsilon_r)+\frac{1}{2}\sqrt{\big( \epsilon_r\big)^{2}+\big( \epsilon_i\big)^{2}}}
\end{eqnarray}
The complex wave propagation vector describes microwave propagation in plasma. By assuming a Y-directed microwave that is propagating in the plasma, the E-field can be written as,
\begin{eqnarray}
    \boldsymbol{E(x,t)}= Re\{E_{0,y} exp (j(\omega t - \boldsymbol{k.}\:x))\}
\end{eqnarray}
$k$ stores the information of the plasma medium in the amplitude and the phase of the wave, based on the $\epsilon(\omega)$ value.  
At critical density ($n_{critical}=\big(m\:\epsilon_0\:\omega_p^{2})/e^{2}$), when the $\omega_p \approx \omega$, the microwave wave starts getting reflected. 
The real part of the Reflection coefficient ($\Gamma_r$) for the plasma and the real part of dielectric permittivity ($\epsilon_{r}$)
 can be related using the following relation,
 \begin{eqnarray} \label{eq1}
    \Gamma_{r}=\bigg( 
    \frac{\sqrt{\epsilon_{r}}-1}{1+\sqrt{\epsilon_{r}}}
    \bigg)
\end{eqnarray}
We can observe from Fig.\ref{Fig:Eprandref} (a) and (b), the Reflection coefficient is high for both collisional and non-collisional plasma for $\omega_p/\omega > 10$ corresponding to $n_{e}$ higher than the $n_{cutoff}=3.84e16$
m$^{-3}$ (where, $n_{cutoff}= n_{critical}(1+(\nu_m/\omega)^{2})$) for a fixed microwave frequency ($f=500$ MHz). Subsequently, there is a drop in the transmission coefficient for the given $\omega_p/\omega > 10$. Also, it is important to note that the $\nu_m$ plays an important role in deciding the cut-off density for a given wave frequency. This can be observed in terms of the broadening of the transmission coefficient curve as well as the reflection coefficient curve as $\omega_p/\omega>10$ instead of steep rise/fall for no-collision case both in reflection and transmission coefficient, respectively. Further, for a collisional plasma, if the plasma density approaches the cutoff density, the plasma shields the incoming microwave resulting in minimum skin depth of EM wave into plasma. 
Based on the different EM wave propagation quantities discussed above, we can say that the plasma density profile controls the different electrical properties of the plasma ranging from dielectric to a conductor. Depending on the relations $\omega > \omega_{p}$, $\omega \approx \omega_{p}$ or $\omega < \omega_{p}$, we can classify the different plasma density regimes as sub-critical, intermediate and over-critical, respectively. Each of those regimes corresponds to transmission, absorption, reflection, and minimum penetration into the plasma, also referred to as skin depth. Thus, by capturing the root mean square (RMS) value of the scattered microwave E-field, we can get the total information regarding the wave propagation in the medium.
\\
\begin{figure}[!htbp]
\centering
\includegraphics[scale=0.35]{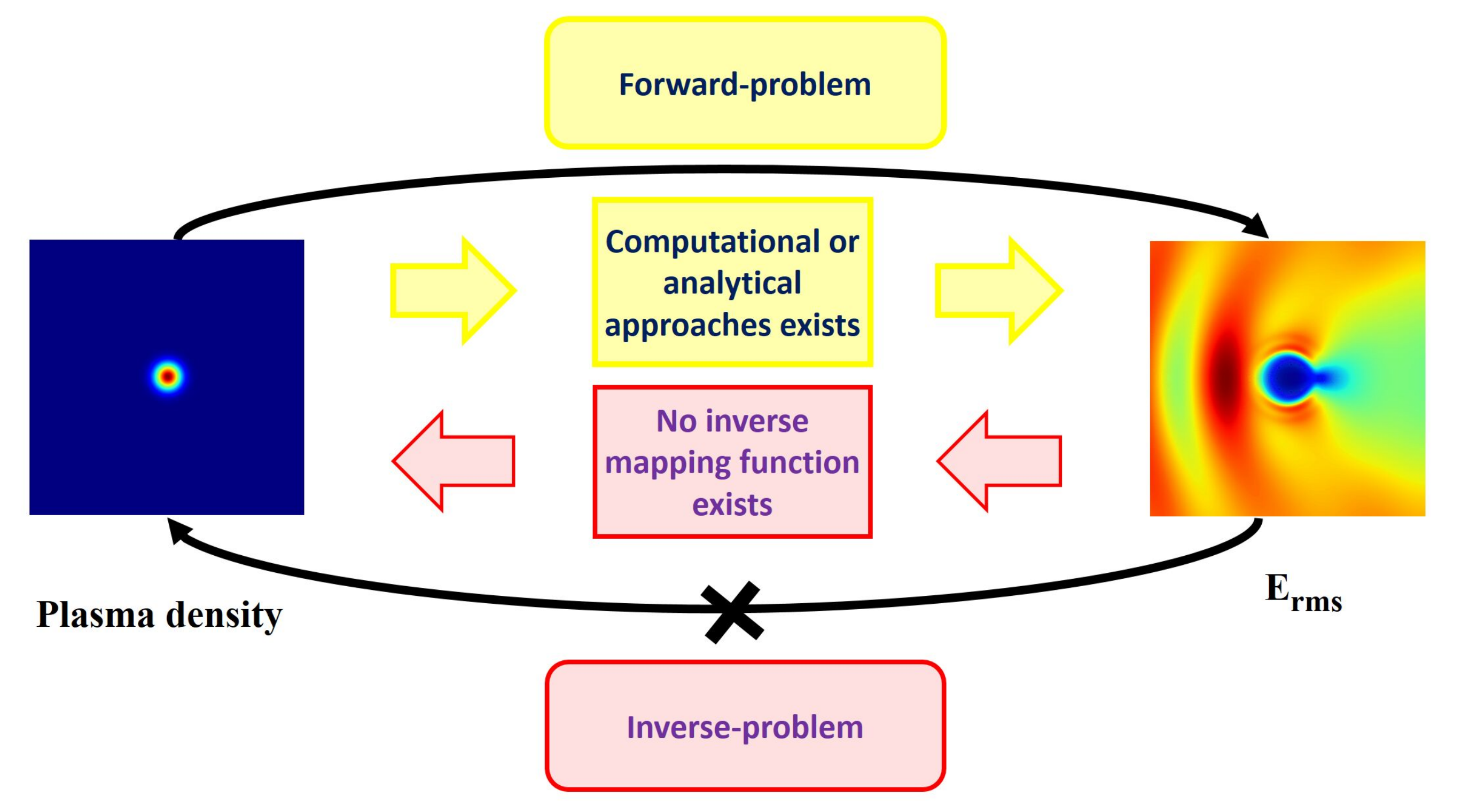}~
\caption{Schematic representation for the category of problems that mostly exists for the EM-plasma interaction. The forward and inverse problem. The 2-D representation of plasma density and corresponding $E_{rms}$ obtained through the Maxwell-plasma fluid model (solution to forward problem) exists is shown, and no direct inverse mapping exists.}\label{Fig:inverseprob}
\end{figure}
\indent
The study of EM wave propagation through plasma is mathematically a well-posed problem, having a unique solution. The problem can be solved computationally by numerical simulations through several techniques, such as FDTD-based iterative solutions of Maxwell-plasma fluid \cite{BHASKAR2007,PGhoshDMR2022,RJVidma1990,AGhayakloo2014,ENoori2022,SZhang2006,GChang2010}. This is referred to as a forward approach (or forward problem). The plasma characterization via wave interaction is a mathematically an ill-posed problem, i.e., given a $E_{rms}$ scattered pattern signature from a plasma density ($n_{e}$) profile, it is difficult to provide information regarding the plasma density profile that has resulted in such pattern. Thus, no unique solution is possible. Such an ill-posed problem to map the plasma density profile for a given electromagnetic field pattern is not directly possible. This is shown by a schematic representation in Fig.\ref{Fig:inverseprob}.
This article proposes an efficient method for $n_{e}$ estimation of plasma density via microwave-plasma interaction-based technique by utilizing the collected scattered electric field pattern measured experimentally and fed to a deep learning-enabled model for the realization of the $n_e$ profile. 
The scattered electric field is to be measured by a sensor network in a plane vertical to the linear plasma device. The sensor deployment varies in terms of sparseness, the distance between individual sensors, the total number of sensors, as well as sensor size. These sensor deployment features majorly depend on the device-specific space availability, along with that how much sparseness is required to produce acceptable results. The sparseness can be random or under favorable conditions can be homogeneous. 
     Generally, electric field induces voltages, and measuring these induced voltages gives the experimental realization of the electric field present within the experimental space. The small dipole antenna (SDA) or parallel plate (PP) type arrangements are some of the easiest ways to measure such induced voltages.  
    For any given electric field the sensor size and the frequency ($\omega$) are the two key parameters that will decide the signal-to-noise ratio (SNR) for the experimental setup. The voltage generated by SDA and PP also strongly depends on the Electric field direction. SDA performs well when placed vertically to the electric field, whereas for PP horizontal direction is the most suitable direction. Therefore the orientation is also important in the sensor deployment \cite{Bassen1983,Gunbooklee2021}. The proposed method works with the measurements of $Erms$ at different sensor locations. The $E_{rms}$ can be conveniently estimated from the induced voltages by taking the RMS value.

\setlength{\textfloatsep}{10pt plus 1.0pt minus 2.0pt}

\section{Data-set Generation Methodology
}\label{sec3}

Sufficient and good quality data in the form of plasma density and the associated scattered electric field when a non-ionizing high-frequency EM wave is incident on the plasma is required for robust application of the proposed deep learning-based technique. The required data has been generated using an in-house Finite-Difference Time-Domain (FDTD) based model, which can capture the EM plasma-interaction in the case of a weakly-ionized, collisional, unmagnetized plasma~\cite{PGhoshDMR2022}. 
When a high-frequency EM wave interacts with such a plasma, the electrons, due to their lower mass, immediately respond to the wave and suffer multiple collisions with the neutrals in every wave cycle. The electrons acquire drift velocity from the wave E-field and subsequently gain instantaneous momentum. However, the high collisions with neutrals interrupt the electron motion, and the electron losses the entire momentum at each collision. The energy exchange between the electrons and neutrals is of the order of ($(m_e/m_{neutral})KE_0$), where $KE_0$ is electron kinetic energy before collision. Since $m_e<<m_{neutral}$, energy remains conserved before and after the collision. Thus, only momentum transfer through collisions dominates and since the EM-wave interaction is non-ionizing, its effect on plasma discharge dynamics is negligible. In the time scale of the microwave, of the order of a few wave periods of a high-frequency wave (say 500 MHz to a few GHz), the plasma density evolution is negligible; therefore, plasma can be considered as a stationary plasma.
 Generally, an FDTD-based model consisting of equations (\ref{eq6}-\ref{eq8}) is adopted for solving the EM-plasma interaction problems in time-domain \cite{PGhoshDMR2022}.
\begin{eqnarray}
\label{eq6}
 \frac{\partial \boldsymbol{E}}{\partial t} \:=\: \frac{1}{\epsilon_0}(\boldsymbol{\nabla}\times \boldsymbol{H})\:-\: \frac{1}{\epsilon_0}\:(\boldsymbol{J})
\end{eqnarray}
 \begin{eqnarray}
\label{eq7}
    \frac{\partial \boldsymbol{H}}{\partial t}\:=\: -\frac{1}{\mu_0}(\boldsymbol{\nabla}\times \boldsymbol{E})
\end{eqnarray}
\begin{eqnarray}
 \label{eq8}
    \frac{\partial \boldsymbol{v_e}}{\partial t}\:=\:-\ \frac{e\ \boldsymbol{E}}{m_e}\:-\:\nu_m\ \boldsymbol{v_e}
 \end{eqnarray}
The model primarily comprises Maxwell's equations and a momentum conservation equation to determine the velocity of an electron. It considers  electric ($\boldsymbol{E}$) and magnetic fields ($\boldsymbol{H}$), permeability ($\mu_0$), electrical permittivity ($\epsilon_0$), electron charge ($e$), electron density ($n_e$), electron velocity ($\boldsymbol{v_e}$), electron mass ($m_{e}$), and electron-neutral collision frequency ($\nu_{m}$) for the solution of the above equations. 
The electron current density is given by, $\boldsymbol{J}=n_eev_e$ in (A m$^{-2}$). The details of the model and its computational implementation can be found in \cite{PGhoshDMR2022,Mihir2022etal}.\\ 
\begin{figure}[!htbp]
\centering
\vspace{-3mm}
\includegraphics[width=\textwidth]{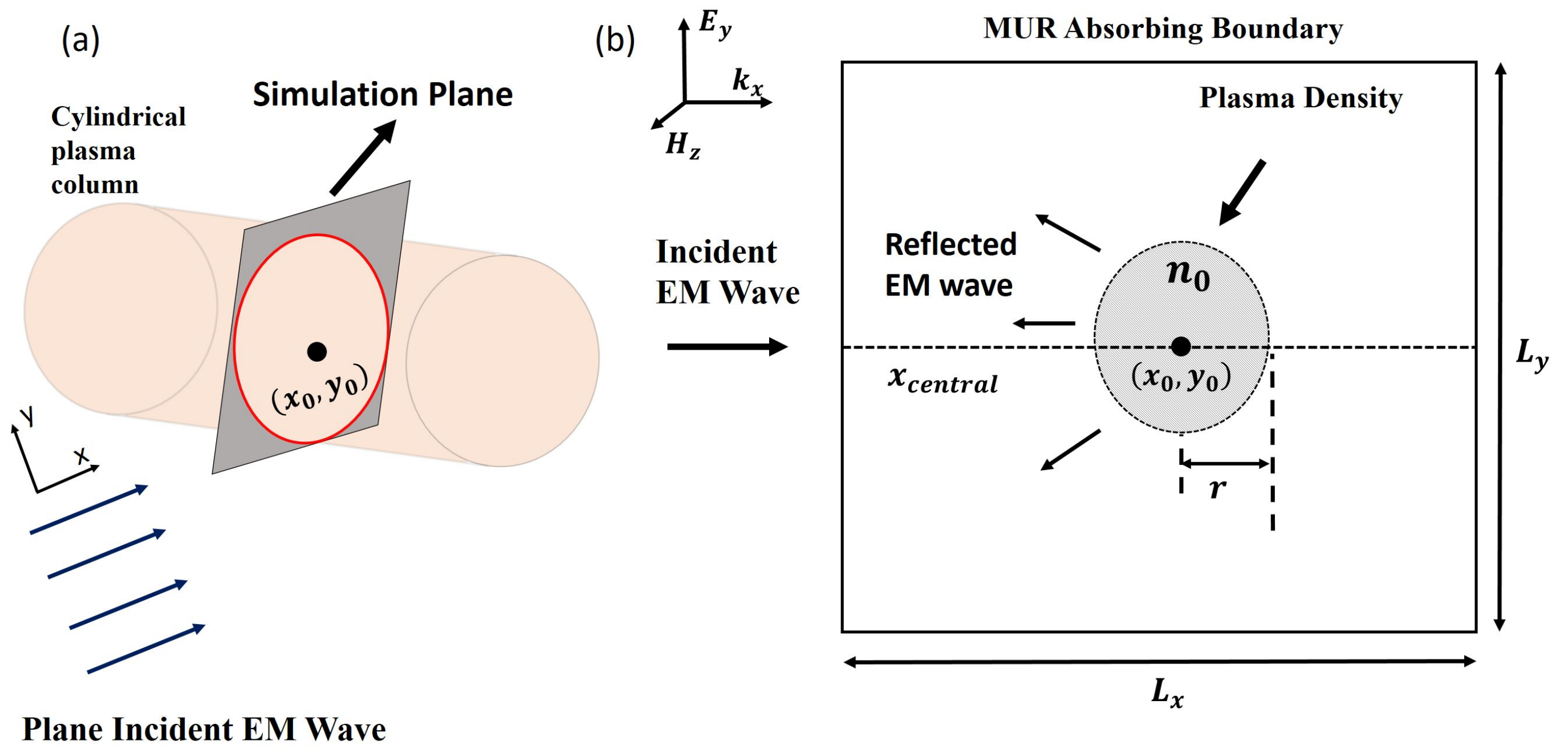}~
\caption{(a). The schematic of a typical linear LTP device. (b). The schematic representation of a square computational domain, the length of the domain $L_{x}$ and $L_{y}$, is expressed in terms of the wavelength of the incident EM wave. The location coordinates $x_{0}$, $y_{0}$ is $0.5L_{x}$ and $0.5L_{y}$, respectively, where $L_x=L_y=1\lambda$, where $\lambda$ corresponds to the freq = 500 MHz. $r$: radius of the plasma and $n_{0}$: peak plasma density .}
\label{Fig:gaussianwidth}
\end{figure}
\indent
Fig.\ref{Fig:gaussianwidth}(b) provides a 
schematic representation of the 2-D simulation domain, where a plane EM wave interacts with a plasma having a cross-sectional radius ($r$) centered at ($x_0,y_0$). The peak density of the plasma is considered as $n_{0}$.
The electric ($E$) field is parallel to the plane of the simulation domain (here XY-plane), and the wave propagation vector ($k$) is parallel to the X-direction. The magnetic ($H$) field is Z-directed, which is perpendicular to the plane of the simulation domain. In the simulation, we have considered a plane EM wave with an amplitude of $10$ V/m and frequency $500$ MHz incident from the left-hand side of the domain as shown in Fig. \ref{Fig:gaussianwidth}. We have considered an air plasma at 2 torr pressure having a collision frequency of around $10$ GHz (for air plasma considered here, $\nu_m=5.3\times10^{9}\:p$, where $p$ is the ambient pressure in (torr))\cite{MYang2021MWreflectpaper}.
Yee-approximation \cite{Yee-1966} has been used to discretize the computational domain of size $1\lambda \times 1\lambda$, where the $\lambda$ corresponds to the EM-wave free-space wavelength. The number of grid points per wavelength of the EM wave is chosen to be 256 to accurately resolve the gradients in the E-field and the plasma density\cite{PGhoshDMR2022}. The resulting total number of grid points in the XY plane is $256\times256$. The data-set size used for training/testing the DL model is equal to the total number of grid points.
The stable scattering pattern is obtained after the EM wave has attained a steady state condition. The simulation model takes the 2D plasma density profile and the incident EM wave properties as inputs and provides the 2-D scattered $E_{rms}$ pattern as outputs. \\
\subsection{Synthetic Data-set preparation for training DL-model}
The data-set to train the network is prepared by keeping the incident wave frequency fixed at 500 MHz and taking different 2D plasma profiles into account. Plasma profiles include symmetric Gaussian and asymmetric non-Gaussian, where the shape and plasma peak density varies. 
The investigations have been carried out in two phases, Phase-1 is associated with symmetric Gaussian Profiles and asymmetric profiles are considered in Phase-2. We briefly discuss the data-set preparation for the Phase-1 experiments, and similar steps are followed for Phase-2.
Different real-life experimental considerations need to be taken into account while preparing the training data-set for the feasible application of the DL-based approach. We have considered the following three possibilities:
\begin{itemize}
   \item The unavailability of the scattered electromagnetic field data within the chamber that confines the plasma.
   \item The collection of either dense or sparse $E_{rms}$ data around the plasma by experimental techniques \cite{M.Nishiuara2017,MAbdurRazzak209,KeeganOrr2020} based on scope of sensor deployment.
   \item Presence of noise in the collected $E_{rms}$ data.  
\end{itemize}
\begin{figure*}[!htbp]
\centering
\includegraphics[width=\textwidth]{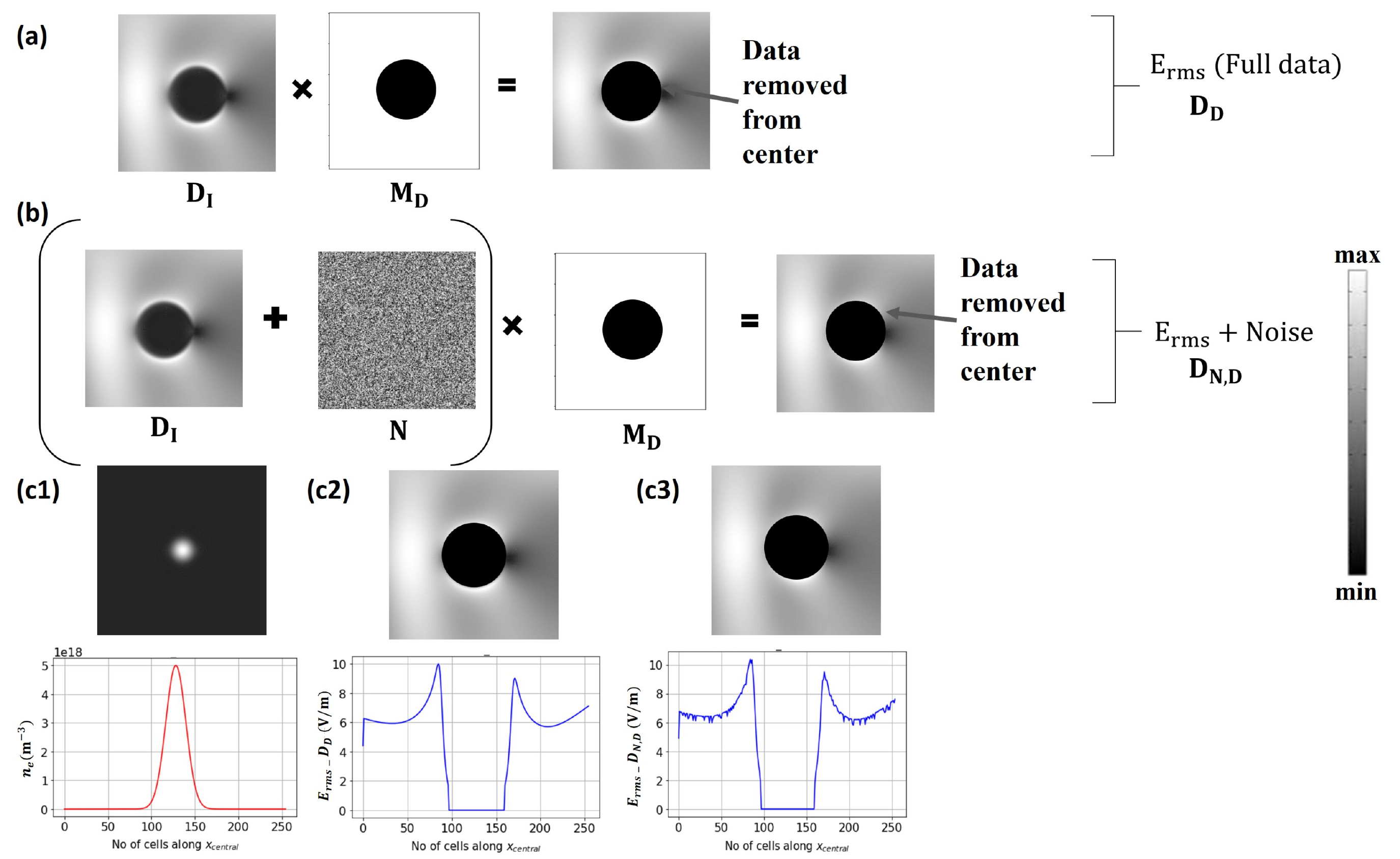}~
\caption{Dense $E_{rms}$ data generation for ML training: (a) by removing the central part of the data and retaining the remaining, (b) addition of noise to the generated dense data, followed by removing the central part and retaining the remaining. (c1-c3): represents the 2-D as well as 1-D density profile, 
and corresponding, dense $E_{rms}$ data collected for both with and without noise (left to right). The color-bar maxima and minima correspond to $E_{rms}$. 
$D_{I}$: Initial data, $M_{D}$: Mask for dense data, $N$: Noise data}\label{Fig:datasetgendense}
\end{figure*}
\begin{figure*}[!htbp]
\centering
\includegraphics[width = \textwidth]{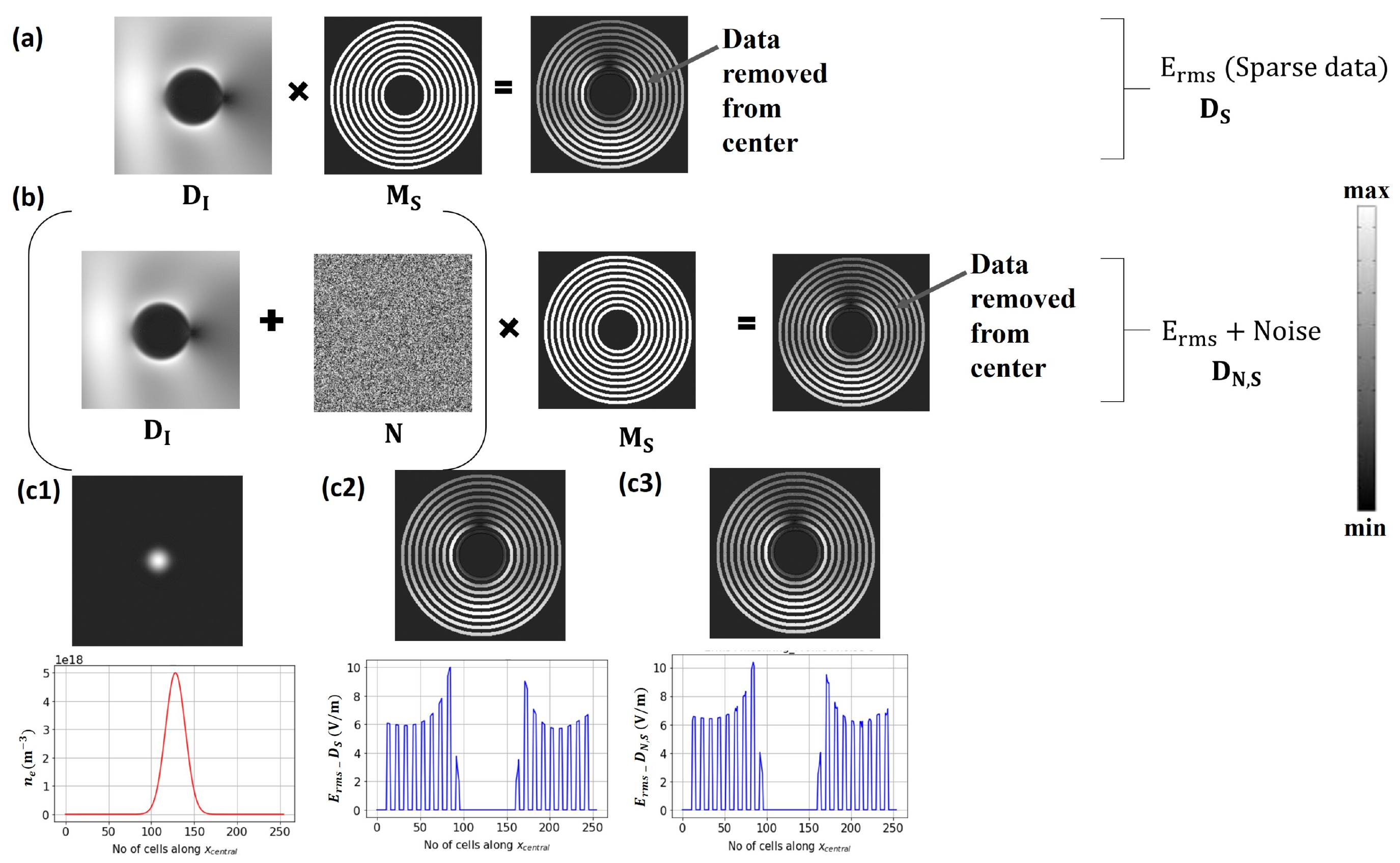}~
\caption{Sparse $E_{rms}$ data generation for ML training: (a) by removing the central part of the data and retaining the sparse data using a concentric ring-based mask, (b) addition of noise to the generated dense data, followed by removing the central part and retaining the remaining. (c1-c3): represents the 2-D as well as 1-D density profile, 
and corresponding, dense $E_{rms}$ data collected for both with and without noise (left to right). The color-bar maxima and minima correspond to $E_{rms}$. 
$D_{I}$: Initial data, $M_{S}$: Mask for sparse data, $N$: Noise data}\label{Fig:datasetgensparse}
\end{figure*}

The unavailability of the data has been considered by masking the central region in the $E_{rms}$ data where we wish to predict the plasma density. The dense or sparse data can be obtained by two different types of masks as shown in figure Fig.\ref{Fig:datasetgendense}, \ref{Fig:datasetgensparse} (a). Applying two different masks leads to either masked dense data, $D_{D}$, or masked sparse data, $D_{S}$.
A mask is equivalent to a non-invasive diagnostic system that records the  $E_{rms}$ data outside the plasma confinement. Here, we have considered the white Gaussian noise model\cite{Ye2020} to mimic various random processes that add from the natural environment to the experimentally collected data, which is $E_{rms}$ data.
The noise model uses a random function to generate random numbers between 0 and 1, which follow a Gaussian distribution. For each of the $E_{rms}$ data samples, the noise magnitude has been restricted to $1-10\%$ of the highest $E_{rms}$ amplitude of that sample. Subsequently, a mask is applied to the noisy $E_{rms}$ data to generate either dense-noisy data or sparse-noisy data. 
Thus, we have obtained four different kinds of $E_{rms}$ data measurements corresponding to a particular density profile, considering the different possibilities in a real experimental setup. This will lead to the following four different data-sets where information in the central part is absent due to masking: 
\begin{itemize}
\item \textbf{$D_{D}$:} dense $E_{rms}$ data-set without noise for different plasma density profiles. Refer Fig.\ref{Fig:datasetgendense} (a).
\item
\textbf{$D_{N,D}$:} dense $E_{rms}$ data-set with noise for different plasma density profiles. Refer Fig.\ref{Fig:datasetgendense} (b).
\item \textbf{$D_{S}$:} sparse $E_{rms}$ data-set without noise for different plasma density profiles. Refer Fig.\ref{Fig:datasetgensparse} (a).
\item \textbf{$D_{N,S}$:}  sparse $E_{rms}$ data-set with noise for different plasma density profiles. Refer Fig.\ref{Fig:datasetgensparse} (b).
\end{itemize}
\begin{figure}[!htbp]
\centering
\includegraphics[scale=0.65]{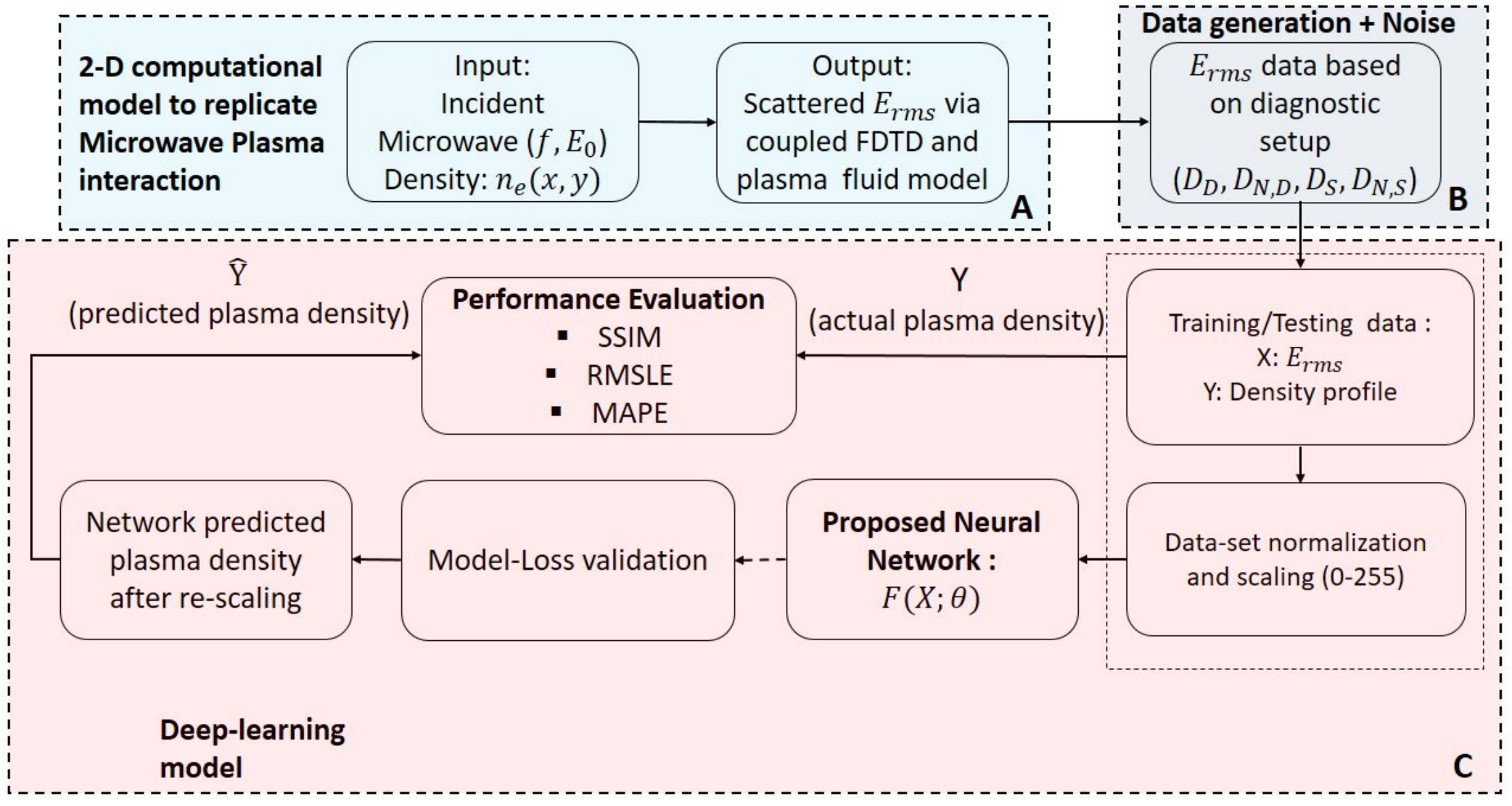}~
\caption{Complete workflow used in this study for prediction of plasma density via DL}\label{Fig:pipeline}
\end{figure}
\section{Deep Learning based Methodology}
The important steps involved in this study, using the proposed data-driven deep-learning (DL) methodology, for the prediction of plasma density from scattered field data obtained outside a plasma chamber are shown in Fig.(\ref{Fig:pipeline}). 
The first step involves the generation of a 2-D scattered $E_{rms}$ field from different plasma profiles for a fixed frequency EM wave incident on the plasma,  using the FDTD-based solver (Figure \ref{Fig:pipeline} - block A). Next, the data processing step is followed (Figure \ref{Fig:pipeline} - block B), which will lead to one of the masked $E_{rms}$ data pattern ($D_{D}$,$D_{S}$,$D_{N,D}$, $D_{N,S}$). The computational steps involved in blocks A and B are explained in the previous section, and we assume that it can also be replicated in a real-experimental diagnostics setup. Training/ testing data consists of a pair of plasma density profiles (Y) and the corresponding masked scattered $E_{rms}$ (X). Figure \ref{Fig:pipeline} - block C shows the use of the suitable DL model (CNN-based UNet described in the next section) followed by training and evaluation. The DL model requires to be trained on an image data-set. Therefore, first, the generated data pair, the masked $E_{rms}$ (Y) and its corresponding plasma density (X), are normalized between 0 and 1 using the already obtained maximum value from the entire training/testing data-set, comprising of the plasma density and $E_{rms}$ data pair. 
Both the maximum plasma density (in m$^{-3}$) and $E_{rms}$ (in V/m) values are saved to help to reconstruct the actual magnitude of the quantity, which the trained neural network will generate, through the re-scaling process. For training the proposed DL network, the normalized data-set is scaled in the range of (0-255), and gray-scale images are generated, which are stored as 4-D image arrays with each of the indices representing the number of images, their dimension, and channel (gray-scale) information. 
The proposed DL model is trained using the generated pair of image-array data-set of masked $E_{rms}$ (X) and plasma density (Y). Since the gray-scale images are generated by scaling the pixel intensity corresponding to the actual normalized value of the data-set pair, proper care must be taken to avoid losing the value due to rounding-off errors. Hence, the image array is fed to the model to avoid errors instead of directly providing the gray-scale image.\\
\indent 
The model is then tested using the remaining data sample (testing) from the train/test data-set. The trained network receives masked $E_{rms}$ image X as input and generates the predicted plasma density image (denoted by $F(X; \theta)$, where F represents the DL model and $\theta$ is the trained model weight matrix). As previously discussed, the predicted plasma density image is converted to physical density values (in m$^{-3}$) by re-scaling the predicted normalized output using the saved global maximum obtained from the entire data-set. The DL model's predicted plasma density values are then compared to the 2-D computational solver's actual plasma density values for evaluation. \\
\subsection{Deep Learning Architecture
}

\indent
\begin{figure*}[!htbp]
\centering
\includegraphics[width = \textwidth]{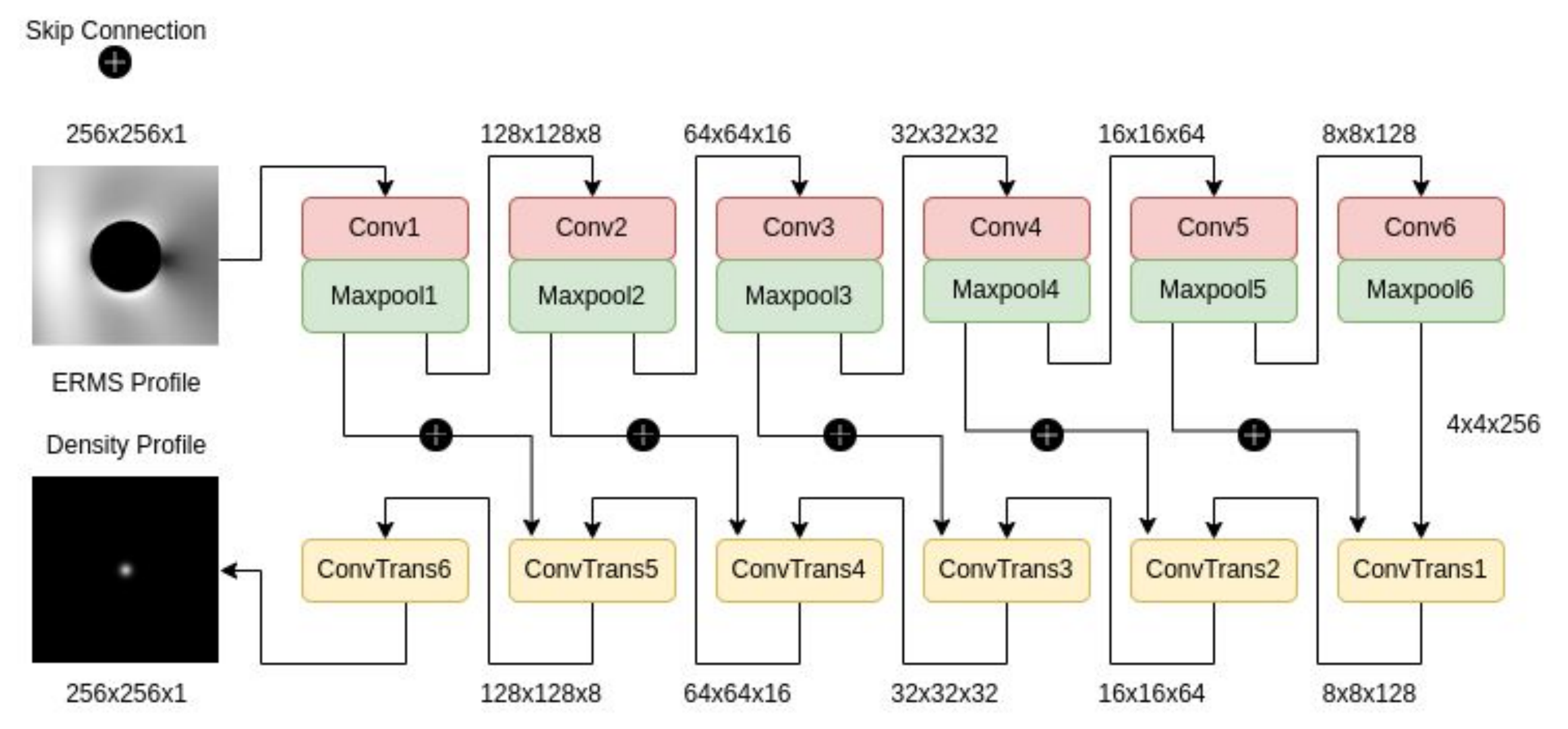}~
\caption{The model architecture uses encoder, decoder and skip connections to predict the plasma density profile from masked EM-wave scattered pattern data-sets}\label{Fig:DL_net}
\end{figure*}
The DL- architecture uses a
CNN-based UNet \cite{b2ORonneberger2015} as depicted in Fig.\ref{Fig:DL_net}. The network consists of two parts, encoders and decoders. The encoder has six convolution layers and six max pool layers. Each convolutional and max pooling layer decreases the input dimension in the encoder to extract finer information at each level. Each of the six encoder units has a convolutional layer with a different number of filters, where the number of filters doubles from the $3\times3$ kernel size of the preceding unit. The ReLU activation function follows the output of each layer. 
There are six decoder units, each with a transposed convolution operation layer with a $2\times2$ kernel size and a ReLU activation function. The decoder layer will upsample the features to create the network output. The output of the encoder is directly coupled to the input of the decoder. The proposed architecture also implements skip connections by connecting the output of each encoder layer to its matching decoder layer, as shown in Fig. \ref{Fig:DL_net}. It helps to solve the vanishing gradient problem by instantly passing the information through the network and propagating it from the shallower layers to the deeper ones. The final decoder unit produces the predicted density profile data based on the masked $E_{RMS}$ data provided as an input.

\setlength{\textfloatsep}{10pt plus 1.0pt minus 2.0pt}

\subsection{Performance metrics}
Performance evaluation of the DL-based approach has been carried out in two steps. Firstly, the images (actual and predicted) of the plasma density ($n_{e}$) are compared using the Structural Similarity Index Metric (SSIM). 
The SSIM~\cite{SSIM} 
is an important metric used to compare the quality of the reconstructed images, where the image degradation is recognized as a change in structural information. 
For two images $x$ and $y$ of size $N \times N$ pixels, the similarity measure is given by:
\begin{eqnarray}
    SSIM = \frac{(2 \mu_x \mu_y + (k_1L)^2)(2\sigma_{xy} + (k_2L)^2)}{(\mu_x^2 + \mu_y^2 + (k_1L)^2)(\sigma_x^2 + \sigma_y^2 + (k_2L)^2)}
\end{eqnarray}
where $\mu$ denotes the mean, $\sigma^2$ denotes the variance, $L$ is the dynamic range of pixel values and $k_1 = 0.01$ and $k_2 = 0.03$. The values of SSIM index lie between 0 and 1. SSIM metric close to 1 suggests good quality of image reconstruction.
\par
To evaluate the performance of the model in predicting the values of $n_{e}$ (in m$^{-3}$) compared to the actual values considered for 2-D FDTD-based fluid-solver, we use two metrics - the RMSLE and MAPE errors metric. The first metric is the average of the absolute percentage error over all the $n_{e}$ values on the 2-D grid. Let $A_{ij}$ and $B_{ij}$ denote the $n_{e}$ values used in the FDTD-based computational solver and $n_{e}$  obtained from DL based approach at $(i,j)^{th}$ point on a $N \times N$ 2D grid, respectively. For the given problem, we have considered only those grid points that lie within a circular cross-section of the plasma confinement (which corresponds to the central region). The mean absolute percentage error (MAPE) is given by
\begin{eqnarray}
    \mbox{MAPE} = \frac{1}{N_{images}}\mathlarger{\sum}_{N_{images}}\left[\frac{1}{N^2}\mathlarger{\sum}_{i=1}^N \mathlarger{\sum}_{j=1}^N \Bigg|\frac{B_{ij} - A_{ij}}{A_{ij}}\Bigg|\right]
\end{eqnarray}
\par

The second metric is the root mean squared logarithmic error (RMSLE) \cite{AAMir2021RMSLE}. The RMSLE is defined as 
\begin{eqnarray}
  RMSLE = \sqrt{\frac{1}{N_{images}} \mathlarger{\sum}_{N_{images}}\left[\frac{1}{N^2} \mathlarger{\sum}_{i=1}^N \mathlarger{\sum}_{j=1}^N (ln(B_{ij}+1) - ln(A_{ij}+1))^2\right]}
\end{eqnarray}

The quantity RMSLE is a better performance metric than the root-mean-square error (RMSE) for the studied problem due to two major reasons. Firstly, high magnitude density data, in both the actual and predicted values, results in a very large RMSE. Secondly, RMSE cannot handle exploding error terms due to outliers that RMSLE can easily scale down and nullify the effects of the prediction error.\\
\section{Results and Discussion}
In this section, we discuss the training details, computational experiments, performance evaluation, and important results obtained from Phase-1 study (symmetric Gaussian profile that uses the dense data ($D_D$, $D_{N,D}$) and the sparse data ($D_S$, $D_{N,S}$)), and Phase-2 study (asymmetric plasma profile where both dense and sparse data has been considered, however, the amount of sparseness has been increased).

\subsection{Training Details}
The deep learning network is separately trained on different data-sets ($D_{D}$, $D_{S}$, $D_{N,D}$ and $D_{N,S}$). The network takes a pair of normalized data matrices, the masked $E_{rms}$, which is given as input to the network (or X) and the corresponding plasma density (Y). The network learns its parameters by minimizing the loss between the actual plasma density (Y) and the output of the network, the predicted plasma density denoted by $F(X; \theta)$ in Fig. \ref{Fig:pipeline}. The loss function for the training of the architecture is given as follows:
\begin{eqnarray}
    L(\theta) = \frac{1}{M} \mathlarger{\sum}_{i=1}^M || F(X;\theta) - Y ||_{2}^{2} + \lambda \mathlarger{\sum}_{j=1}^l ||W_j||_{1}
\end{eqnarray}
where $M$ is the total number of training samples, $\theta$ represents the network weight parameter matrix, $l$ is the total number of kernels used and $W_{j}$ is the weight of the $j^{th}$ kernel. The optimizer used for the training is Adam optimizer \cite{Kingma} with a learning rate of $\alpha = 0.001$ and $\epsilon=10^{-7}$ for numerical stability. The exponential decay for the first moment has been taken as $\beta_{1}=0.9$, and the exponential decay for the second moment is $\beta_{2}=0.999$. L1 regularization is used to counter the problem of overfitting with $\lambda = 10^{-6}$. Uniform Xavier initialization \cite{Glorotuniform2010} is used as the kernel initializer. The normalized initialization of the Weights of each layer can be heuristically expressed as, 
\begin{equation}
W \sim U\left[ - \frac{\sqrt{6}}{\sqrt{n_{j} + n_{j+1}}}, \frac{\sqrt{6}}{\sqrt{n_{j} + n_{j+1}}} \right]
\end{equation}
where, $U$ is the uniform distribution and $n_j$ is the number of nodes in $j^{th}$ layer. The proposed deep learning model, consisting of six convolutional layers, six Transpose convolution layers, and 5 skip connections, has 611,833 trainable parameters. The network is trained on NVIDIA Tesla K40c GPU using Keras API with TensorFlow running in the backend.\\ 
\subsection{Phase-1 : Experiments with dense data ($D_{D}$ and $D_{N,D}$)}
\begin{figure*}[htbp]
    \centering
       \includegraphics[width=\textwidth]{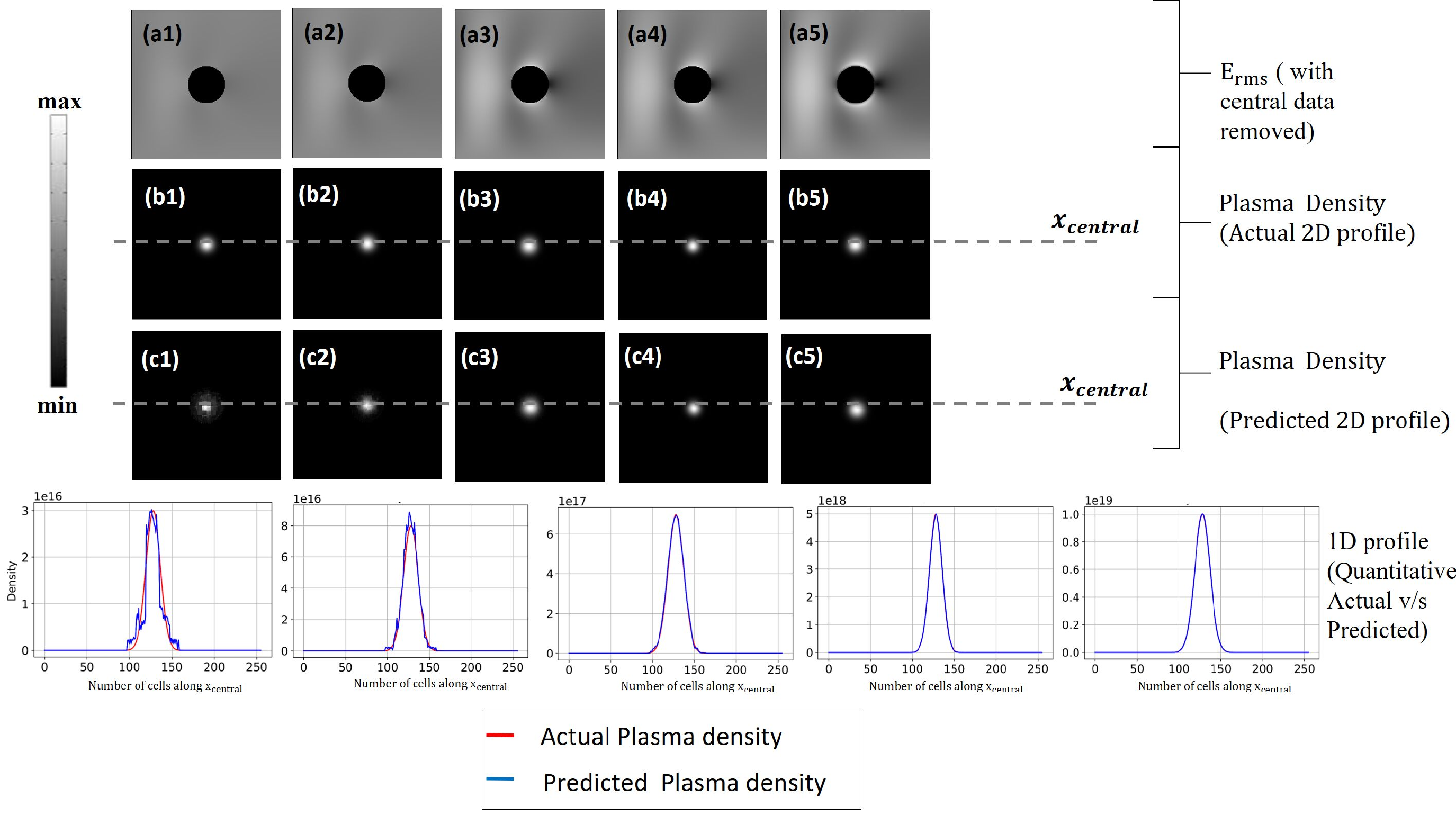}
        \caption{Comparative study and results for dense data-set ($D_{D}$); Row 1 (a1-a5): 
        masked 2-D $E_{rms}$  dense data obtained from simulations for given input density profile; Row 2 (b1-b5): The actual input plasma 2-D density profile; Row 3 (c1-c5): The predicted 2-D profile of plasma density from the proposed deep learning based architecture;  Row 4: Comparison of the accuracy between the magnitude of the actual and predicted 1-D 
        plasma density along the central x-axis ($x_{central}$) of the computational domain. 
        }
        \label{fig:DLcomparefulldat}
    \end{figure*}

\begin{figure*}[htbp]
    \centering
      \includegraphics[width=\textwidth]{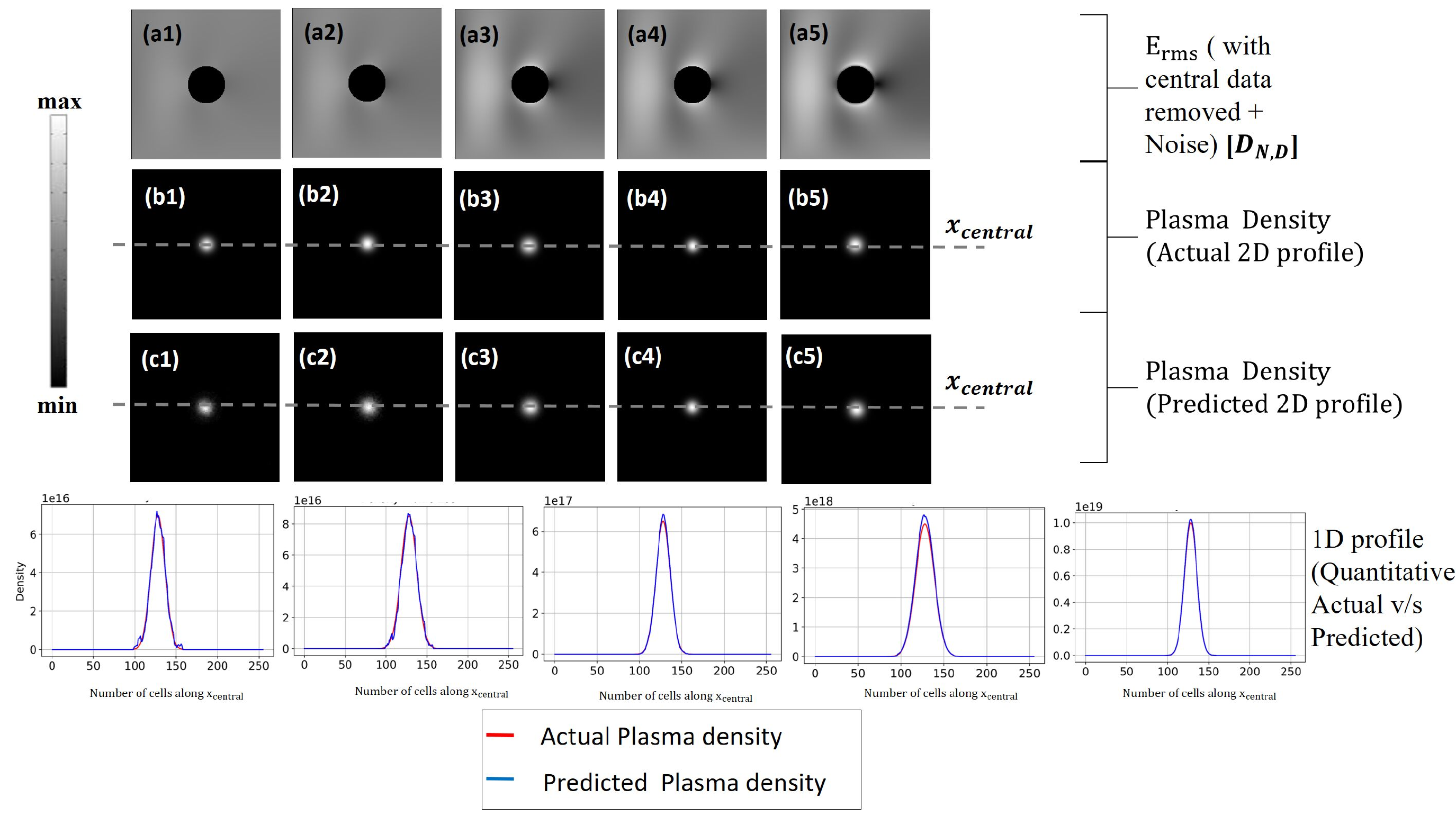}
        \caption{
        Comparative study and results for dense data-set with noise ($D_{N,D}$); Row 1 (a1-a5): 
        2-D $E_{rms}$ (the increasing maxima in V/m from left to right) dense data with Gaussian noise obtained from plasma density profile using FDTD; Row 2 (b1-b5): The actual plasma 2-D density profile; Row 3 (c1-c5): The predicted 2-D profile of plasma density from the proposed deep learning based architecture;  Row 4: Comparison of the accuracy between the magnitude of the actual and predicted 1-D 
        plasma density along the central x-axis ($x_{central}$) of the computational domain in the presence of noise.      
        }
        \label{fig:DLnoisefulldat}
    \end{figure*}

\begin{table}
\centering
\caption{ RMSLE and MAPE-based comparison of predicted plasma density with the actual density data for different types of dense data samples}
  \begin{tabular}{|l|l|l|l|l|l|l|l|}
    \hline
     \multirow{2}{*}{\begin{tabular}[c]{@{}l@{}}Types of\\ Data sample\end{tabular}} &\multicolumn{6}{c|}{ Peak plasma density ($n_{0}$) (m$^{-3}$)}&\multirow{2}{*}{\begin{tabular}[c]{@{}l@{}}SSIM\\ 1e16-1e19\\ (m$^{-3}$)\end{tabular}}\\
       \cline{2-7}
      &\multicolumn{2}{c|}{1e16-1e17}&\multicolumn{2}{c|}{1e17-1e18}&\multicolumn{2}{c|}{1e18-1e19}&\\
     \cline{2-7}
      &\multicolumn{2}{c|}{\begin{tabular}[c]{@{}l@{}}sub-critical\\ $\omega>\omega_p$\end{tabular}}&\multicolumn{2}{c|}{\begin{tabular}[c]{@{}l@{}}intermediate\\ $\omega\approx\omega_p$\end{tabular}}&\multicolumn{2}{c|}{\begin{tabular}[c]{@{}l@{}}over-critical\\ $\omega<\omega_p$\end{tabular}}&\\
      \cline{2-7}
      & RMSLE & MAPE &RMSLE & MAPE & RMSLE & MAPE & \\
    \hline
   $D_{D}$& {0.142} & {0.12} & {0.031}&{0.024}  & {0.025}  &{0.016} & {0.9998}\\
   $D_{N,D}$& 0.166 &0.153  &0.057 & 0.048 &0.043  &0.038  &0.9995 \\
   \hline
  \end{tabular}
  \label{table1a}
\end{table}
We generated the dense dataset without noise ($D_{D}$) by changing the Gaussian plasma density profiles with $n_{0}$ ranging from $1e16$ to $1e19$ m$^{-3}$. The data-set comprised 8000 pairs (density profile and masked scattered $E_{rms}$) of samples.
Subsequently, a dense-dataset with noise ($D_{N,D}$) is prepared as described in section \ref{sec3}.1.
DL model has been trained separately with $D_{D}$ and $D_{N,D}$. Both the data-sets were divided in the ratio of 80 to 20 data-samples for training and testing, and the test data-set is further divided with a similar ratio for cross-validation. The MAPE and RMSLE metrics have been separately reported (in Table \ref{table1a}) for different ranges of $n_{0}$ to understand the prediction capability of the proposed approach in different density ranges. The SSIM metric is reported for the overall range of plasma densities ($1e16-1e19$ m$^{-3}$), and we observe that the overall SSIM is very high ($~0.999$). RMSLE and MAPE is $<0.1$ for $n_{0}$  $>1e17$ m$^{-3}$ for both $D_{D}$ as well as $D_{N,D}$, but $>0.1$ for density range $1e16-1e17$ m$^{-3}$.  
 The better prediction for higher density values can be attributed to the high reflection component in the scattered $E_{rms}$ pattern, which gets more appropriately captured as features by the DL model.  
 We observe improvement in both MAPE and RMSLE (Table \ref{table2}) when the model is trained with samples in the overcritical density range ($1e18-1e19$ m$^{-3}$). We also observe that an acceptable prediction can be obtained even using a surprisingly small Data-set size (less than 1000 measurements).\\
\indent
MAPE, RMSLE, and SSIM provide a single number summary about the predictive capability of the proposed approach; however, to obtain a complete qualitative as well as a quantitative understanding of the predicted values of plasma density, we have performed a 2D data analysis as shown in Fig.\ref{fig:DLcomparefulldat} (for $D_D$) and Fig.\ref{fig:DLnoisefulldat} (for $D_{N,D}$). The first row (a1-a5) in both the figures (Fig. \ref{fig:DLcomparefulldat} and Fig.\ref{fig:DLnoisefulldat}), indicates the masked scattered $E_{rms}$ pattern, the input to the model. The second row (b1-b5) shows the corresponding actual density profile based on which the FDTD-based computational solver generated the scattered $E_{rms}$ pattern. In both, Fig.\ref{fig:DLcomparefulldat} and Fig.\ref{fig:DLnoisefulldat}, plasma density varies from low to high values (b1-b5). Based on the plasma density profile, the $E_{rms}$ pattern varies, indicating transmission, reflection, and absorption of the propagating microwave. 
The predicted 2-D profile of plasma density ($n_{e}$) from the DL network is shown in row 3 (c1-c5). We observe a good qualitative match with the actual density profile in row 2 (b1-b5). In row 4 (Fig. \ref{fig:DLcomparefulldat} and Fig.\ref{fig:DLnoisefulldat}), 1-D comparison between the actual and the predicted $n_{e}$ profile along the central x-axis ($x_{central}$) shows a good quantitative match between the two. We observe better predictions for dense plasma with peak density ($n_{0}$), $> 1e17$ m$^{-3}$. Thus, it validates the observed trend of low MAPE and RMSLE in Table \ref{table1a} and Table \ref{table2}.
We have conducted another set of experiments with samples in the density range $1e18-1e19$ m$^{-3}$ to understand whether training the model with a narrow range of density values leads to better predictive ability. This experiment is also aimed at determining the minimum number of measurements that is required for DL-based prediction with desirable accuracy.
\begin{table}
\centering
\caption{Performance evaluation for computational experiments performed with dense data samples ($D_{D}$) in the limited range having $n_{0}:1e18-1e19$ m$^{-3}$}
  \begin{tabular}{|l|l|l|l|}
    \hline
     \multirow{2}{*}{\begin{tabular}[c]{@{}l@{}}Number of\\ Data samples\end{tabular}} &\multicolumn{3}{c|}{ $n_{0}:$ $1e18-1e19$ (m$^{-3}$)}\\
       \cline{2-4}
       &RMSLE&MAPE&SSIM\\
      \hline
      
      200& 0.042 & 0.032 & 0.9978\\
      500& 0.041 & 0.030 & 0.9987\\
     750& 0.034 & 0.029 & 0.9992\\
     1000& 0.021 & 0.025 &0.9994 \\
     1500& 0.0156 & 0.021 & 0.9996\\
    \hline
  \end{tabular}
  \vspace{2pt}
  \label{table2}
\end{table}
\begin{table}
\centering
\caption{ RMSLE and MAPE-based comparison of predicted plasma density with the actual density data for different sparse data sets}
  \begin{tabular}{|l|l|l|l|l|l|l|l|}
    \hline
     \multirow{2}{*}{\begin{tabular}[c]{@{}l@{}}Types of\\ Data sample\end{tabular}} &\multicolumn{6}{c|}{ $n_{0}$ (m$^{-3}$)}&\multirow{2}{*}{\begin{tabular}[c]{@{}l@{}}SSIM\\ 1e18-1e19\\ (m$^{-3}$)\end{tabular}}\\
       \cline{2-7}
       &\multicolumn{2}{c|}{1e16-1e17}&\multicolumn{2}{c|}{1e17-1e18}&\multicolumn{2}{c|}{1e18-1e19}&\\
      \cline{2-7}
      & RMSLE & MAPE &RMSLE & MAPE & RMSLE & MAPE & \\
     \hline
  $D_{S}$&0.054  & 0.038 & 0.031& 0.023 & 0.012 & 0.010 & 0.9999 \\
   $D_{N,S}$& 0.188  & 0.11 &0.051 & 0.041 &0.035  &  0.027& 0.9996\\
    \hline
 \end{tabular}
 \label{table1b}
\end{table}
\begin{figure*}[htbp]
    \centering
      \includegraphics[width=\textwidth]{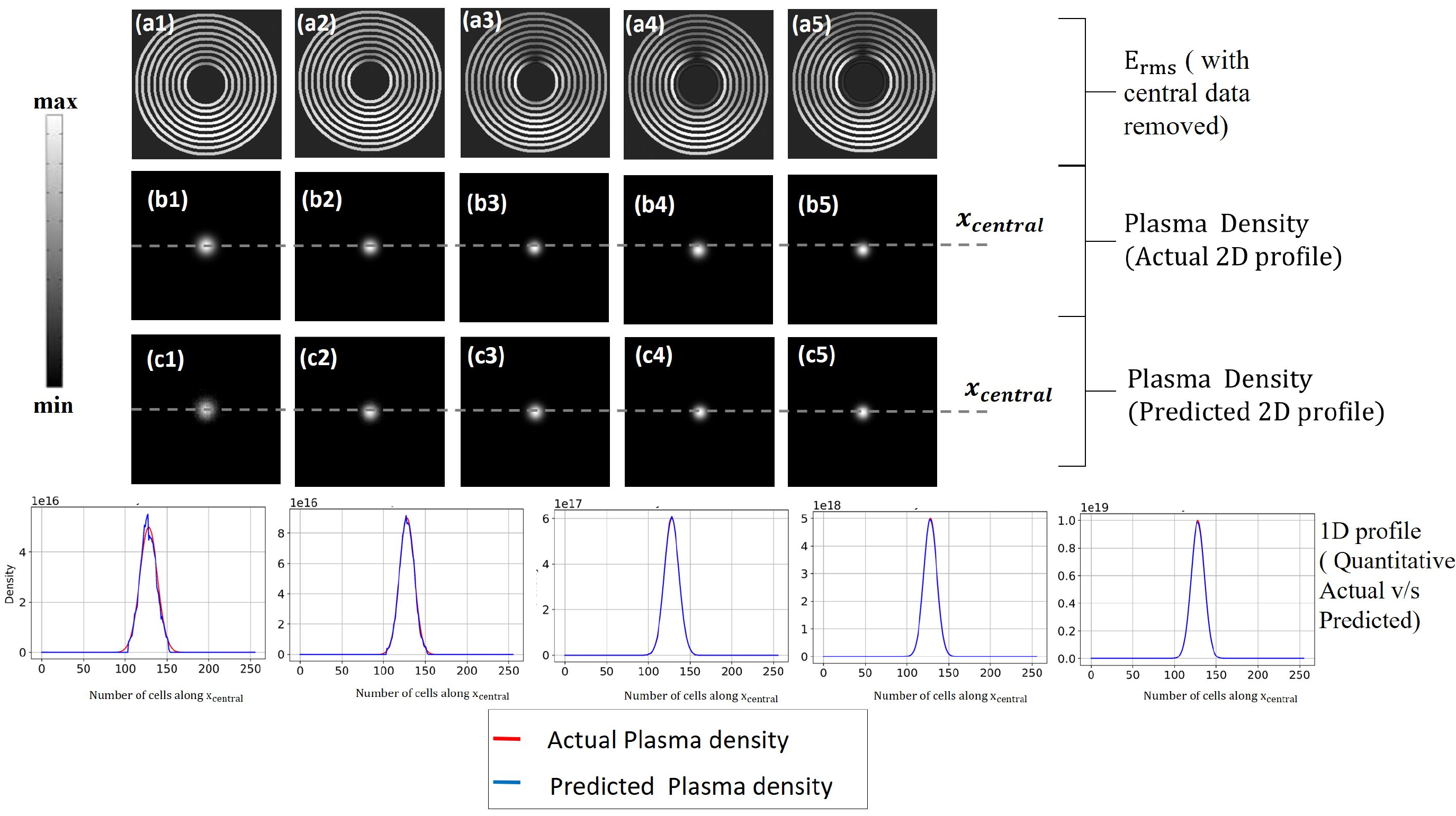}
        \caption{
        Comparative study and results for sparse data-set ($D_{S}$); Row 1 (a1-a5): 
        2-D $E_{rms}$ (the increasing maxima in V/m from left to right) sparse data obtained from plasma density profile using FDTD; Row 2 (b1-b5) The actual plasma 2-D density profile; Row 3 (c1-c5): The predicted 2-D profile of plasma density from the proposed deep learning based architecture;  Row 4: Comparison of the accuracy between the magnitude of the actual and predicted 1-D 
        plasma density along the central x-axis ($x_{central}$) of the computational domain. }
        \label{fig:DLcompareRings}
    \end{figure*}
\begin{figure*}[htbp]
    \centering
       \includegraphics[width=\textwidth]{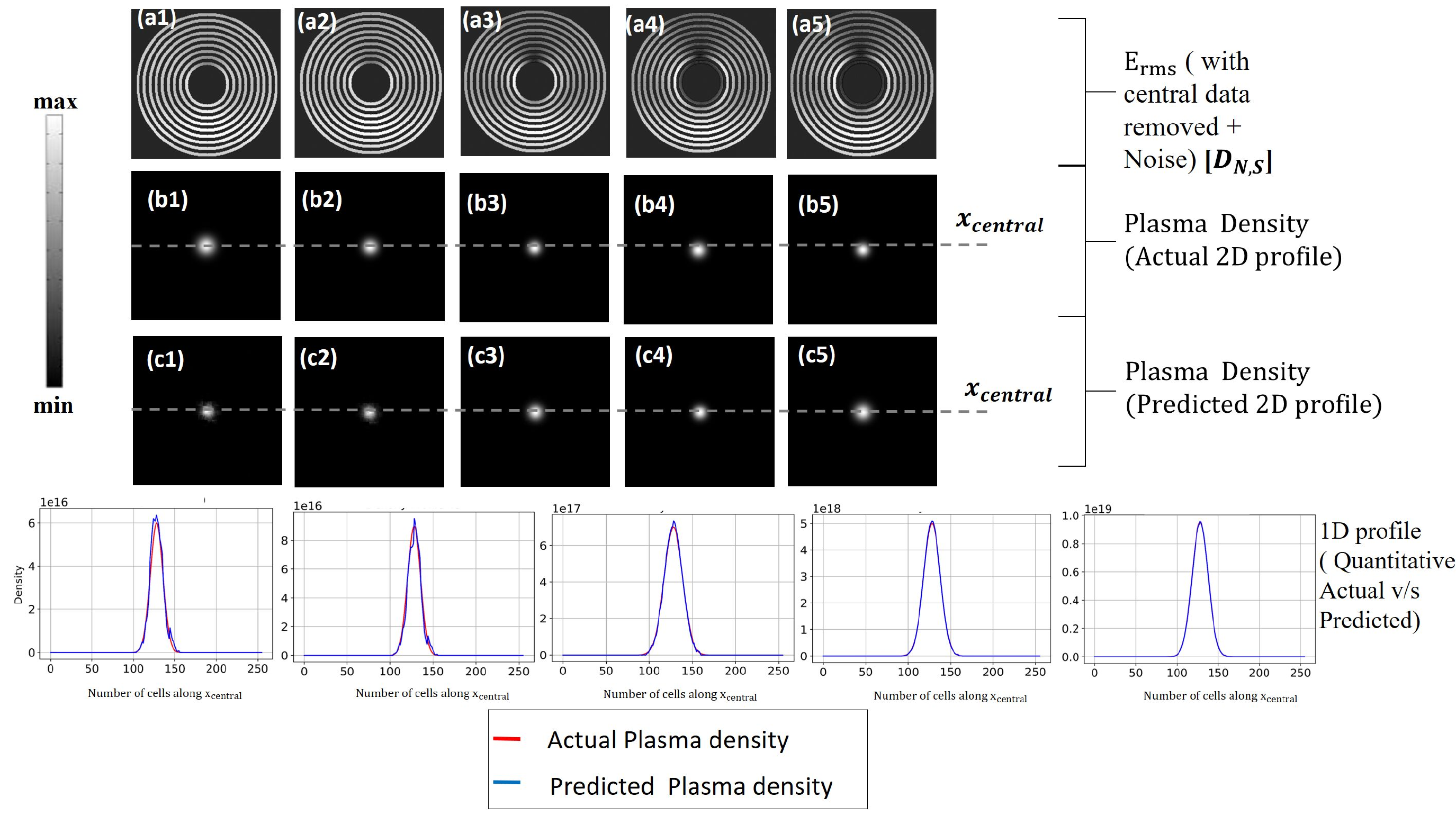}
        \caption{
        Comparative study and results for sparse data-set with noise ($D_{N,S}$); Row 1 (a1-a5): 
        2-D $E_{rms}$ (the increasing maxima in V/m from left to right) sparse data with Gaussian noise obtained from plasma density profile using FDTD; Row 2 (b1-b5) The actual plasma 2-D density profile; Row 3 (c1-c5): The predicted 2-D profile of plasma density from the proposed deep learning based architecture;  Row 4: Comparison of the accuracy between the magnitude of the actual and predicted 1-D 
        plasma density along the central x-axis ($x_{central}$) of the computational domain in the presence of noise.           
        }
        \label{fig:DLnoiseRingdata}
    \end{figure*}
\subsection{Phase-1: Experiments with sparse data ($D_{S}$ and $D_{N,S}$)}    
 A similar study is repeated with the sparse $E_{rms}$ data-set, for both without ($D_{S}$) and with noise ($D_{N,S}$). The data-sets comprised 7000 samples each for without and with noise, with plasma peak density ($n_{0}$) varying in the range $1e16-1e19$ m$^{-3}$. Based on training with the masked sparse $E_{rms}$ - $n_{e}$ data pairs, the trained DL-model has been used to predict the unknown $n_{e}$ for a masked sparse $E_{rms}$ test sample. The error metrics RMSLE and MAPE has been separately reported for different range of $n_{0}$ in Table \ref{table1b}. In addition, the SSIM metric is reported for the overall $n_{0}$ range, which is very high ($~0.999$). Even with sparse data where a significant amount of scattered E-field information is absent, based on the metrics' results in Table \ref{table1b}, we can infer that the proposed approach can determine the plasma density within an acceptable range. 

 \indent
A comparison between the actual and the predicted 2D plasma profiles for five test samples with different peak density values (under-dense (leftmost) to over-dense plasma (rightmost)) have been shown in Fig.\ref{fig:DLcompareRings} (for $D_{S}$) and Fig.\ref{fig:DLnoiseRingdata} (for $D_{N,S}$). We observe a good qualitative match in both cases with better results for a higher range of density values (for the intermediate and overcritical plasma density regime).
1-D comparison between the actual and the predicted $n_{e}$ profile  along the central x-axis (row 4 in Fig.\ref{fig:DLcompareRings} and Fig. 
\ref{fig:DLnoiseRingdata}) shows a good quantitative match.
Our study shows that the proposed methodology can also be employed with good confidence with sparse measurements of scattered E-field signals outside the plasma with a non-invasive approach.  \\
\subsection{Phase-2 : Experiments with asymmetric data}
Asymmetric profiles are the most obvious situations that
may be encountered in real experiments. Phase-2 experiments with asymmetric profiles have been performed in the density range ($1e18-1e19$ m$^{-3}$), where we found more accurate results in Phase-1, due to the dominance of the EM wave reflections, over the transmission or absorption in the scattered $E_{rms}$ pattern (which the DL-model extracts as important features for learning the mapping between input and output). 
Asymmetry has been introduced in the data-sets by multiple ways, firstly by considering Gaussian profiles that are not at the center of the simulation domain ($x_0,y_0$) but located at random locations (top, bottom, left or right relative to $x_0,y_0$ ), secondly by considering two Gaussian profiles and thirdly by considering non-Gaussian profiles with multiple peaks. The first two data-sets are referred here as partial asymmetric, which uses  9000 data samples, while the third data-set as full asymmetric data-set, which uses 1200 data samples. Both dense ($D_{D}$) and sparse ($D_{S}$) data-sets have been considered in phase-2; however, we have designed a more difficult learning and prediction problem by introducing sharp gradients in the asymmetric density profiles and considering more sparse scattered field data in sparse data-set ($D_{S}$).\\
    \begin{figure}[!htbp]
    \centering
    \includegraphics[width=\textwidth]{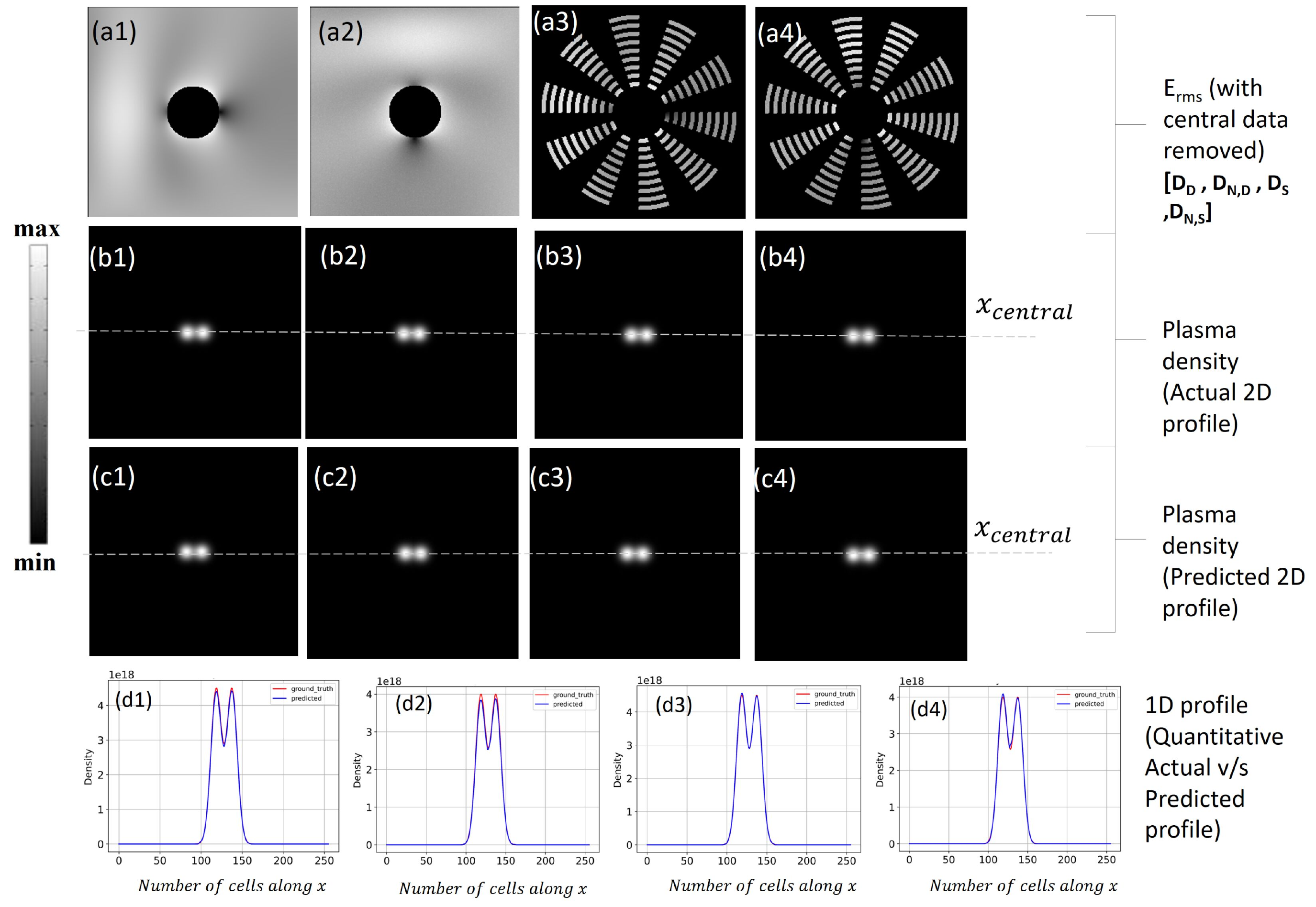}~
    \caption{Comparison of results for dense ($D_D$) and sparse ($D_S$) data samples without and with noise (for partially asymmetric), having two Gaussian plasma profiles with peaks not located at the center of the simulation domain ($x_0,y_0$).
    }\label{Fig:symmetricdense}
    \vspace{-5mm}
    \end{figure}
Comparison between the actual and the predicted 2-D plasma profiles for four test samples (two each for dense and very sparse) for different locations of the peak density values in the case of partial asymmetric data-set have been shown in Figure \ref{Fig:symmetricdense}. We observe a good qualitative match for both dense and very sparse data. 1-D comparison between the actual and the predicted $n_{e}$ profile along the x-axis (row 4 in Fig.\ref{Fig:symmetricdense}) shows a good quantitative match. \\
    \begin{figure}[!htbp]
    \centering
    \includegraphics[width=\textwidth]{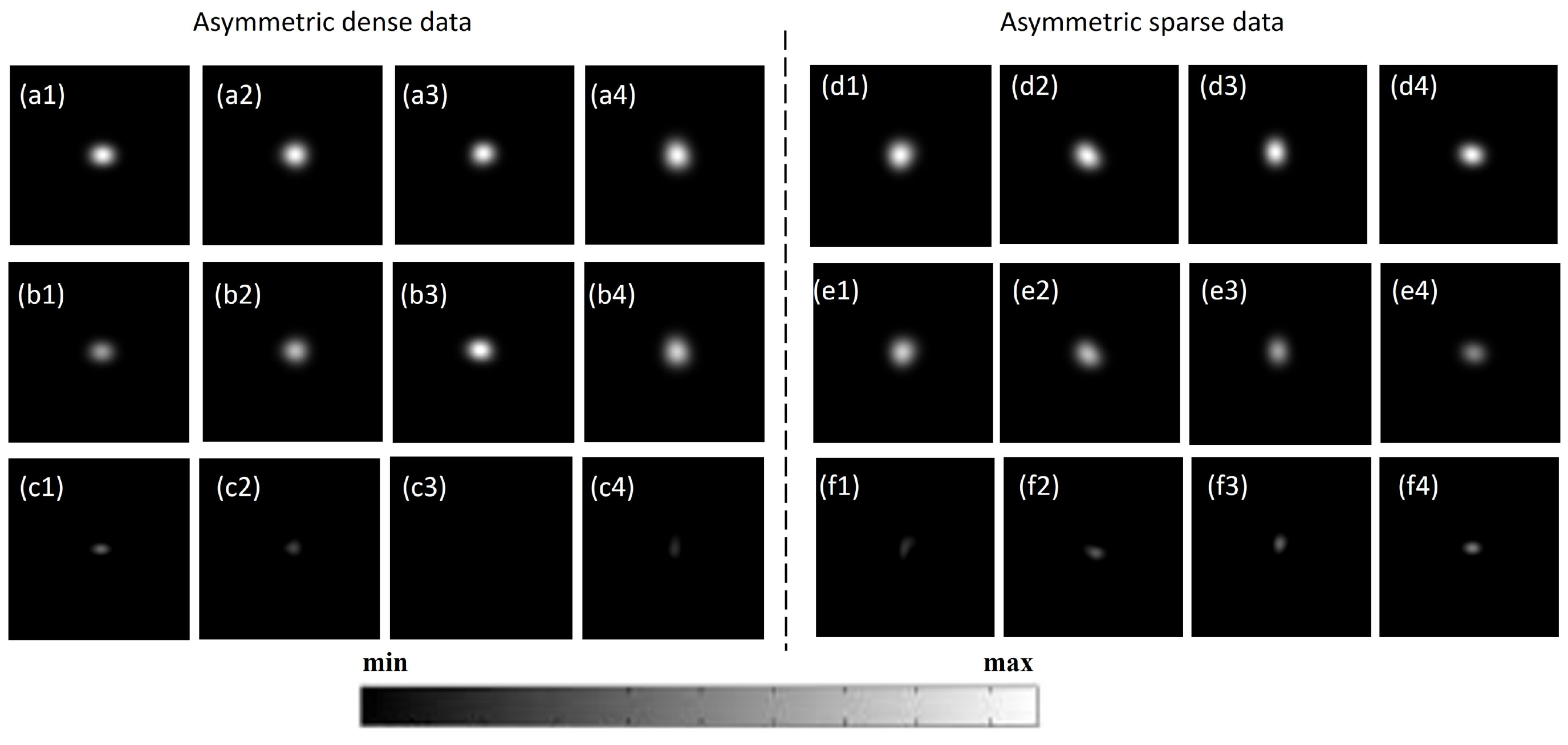}~
    \caption{Comparison of results for completely asymmetric density profile for dense and sparse $E_{rms}$ data, $D_D$ and $D_{S}$, respectively. (a1-a4) the actual, and, (b1-b4) the predicted plasma density profile for $D_D$. 
    (c1-c4) the residual (difference between actual and predicted) to indicate the degree of mismatch. (d1-d4) the actual, and, (e1-e4) the predicted plasma density profile for $D_S$. (f1-f4) the residual to indicate the degree of mismatch between actual and predicted asymmetric profiles. 
    }\label{Fig:asymmetricdense}
    \end{figure}
We use 2-D plots of the actual and model-predicted plasma profile along with the residual to evaluate the results for fully asymmetric data (non-Gaussian shapes with multiple peaks) as shown in Fig. \ref{Fig:asymmetricdense}.
 We can observe that the DL model can capture asymmetry (either multiple peaks with different peak densities that are merged together, locations of the different peaks, or the shape of the profile, such as elliptic, etc.) in the plasma profile within acceptable accuracy. It is interesting to observe that the model-predicted plasma density profile has a good spatial resolution that matches the actual plasma profile, and we obtain an average SSIM of more than .99 for different samples. Also, we can observe that the DL model can preserve the order of the plasma peak densities within a desirable accuracy.
 The decrease in accuracy compared to Phase-1 results (Gaussian profiles) can be attributed to multiple factors, particularly high-density gradients and asymmetry leading to complex scattering patterns. The results indicate that the DL-based approach can predict the profile shape, the location of plasma peaks, and the peak plasma density with desirable accuracy, the basic diagnostic requirement for any real laboratory experiment.  \\
\begin{table}
\centering
\caption{Performance evaluation for computational experiments performed with $E_{rms}$ data both (dense and sparse) corresponds to plasma density profile having partial and full asymmetry. The peak density range having $n_{0}:1e18-1e19$ m$^{-3}$}
  \begin{tabular}{|ll|l|l|l|}
    \hline
     \multicolumn{2}{|c|}{\begin{tabular}[c]{@{}l@{}}Types of density profiles\\ and Data samples\end{tabular}} &\multicolumn{3}{c|}{ $n_{0}:$ $1e18-1e19$ (m$^{-3}$)}\\
       \cline{3-5}
     & &RMSLE&MAPE&SSIM\\
      \hline
       &$D_D$ & 0.03 & 0.024 & 0.9998\\
        Partial asymmetry&&&&\\
      &$D_S$ & 0.0425 & 0.035 & 0.9998\\
     \hline
    &$D_D$ & 0.3& 0.28 & 0.9963\\
     Fully asymmetry&&&&\\
   &$D_{S}$ & 0.24 & 0.18 & 0.9964\\
  \hline
  \end{tabular}
  \vspace{2pt}
  \label{table4}
\end{table}
\setlength{\textfloatsep}{10pt plus 1.0pt minus 2.0pt}
\section{Conclusion}
For the first time, a novel deep learning-based plasma diagnostics approach is presented where the combination of microwave-plasma interaction physics, existing plasma diagnostics techniques, and deep learning to train neural networks for plasma density prediction with high accuracy and minimal efforts are proposed. The  approach is based on computational experiments involving the scattering of microwaves by considering an unmagnetized, collisional , partially ionized low temperature plasma. Experimentally, the $E_{rms}$ generated from the scattering of the microwave are processed for the estimation of $n_e$. The approach is applied to different experimental conditions, a range of noises, multiple density profiles (symmetric and asymmetric), and sparseness of the $E_{rms}$ collection as well as their combinations too. Every condition has experimental significance, addressing the experimental limitations. The SSIM for a different combination of the experimental conditions for the density estimation remains near $\approx 0.99$, which suggests convincing evidence of accurate $n_e$ estimations. The DL model performed well in reproducing the plasma density profile under the different possible experimental conditions. The predicted results are within acceptable ranges provided the profiles are more symmetric and having higher plasma density for both dense and sparse data. The percentage error (MAPE and RMSLE) in predictions lies within 1 to $10\%$. The network performs well even for noisy $E_{rms}$ data and able to predict partially asymmetric plasma profiles both in the presence and absence of noise, the percentage error lies within the similar range of 1 to 10\%. We observed around 20\% increase in the percentage error from the model prediction for asymmetric plasma profile. The presence of sharp gradients in the asymmetric profiles may result in the abrupt $E_{rms}$ pattern which can be a possible reason for error in prediction. Use of a larger training data-set or proper noise filtering techniques  may allow smoothing out the predicted profile, further lowering the percentage error for asymmetric profile prediction. We also plan to explore the use of customized loss functions to further improve the predictions in the case of asymmetric plasma profiles. Currently, the proposed method is being applied to the real experimental data and the performance is under evaluation, the results will be communicated soon. The model will be refined/ modified to handle the fusion grade plasma, tokamak plasma, where the temperature and density are high as well as strong magnetic fields are available. 

\section*{Acknowledgment}
The authors P. Ghosh and B. Chaudhury acknowledge the DST-SERB for financial assistance (Project No. - CRG/2018/003511). The authors would also like to gratefully acknowledge DA-IICT, India for providing the computational facilities and kind support to carry out this research work.
\bibliography{references}

\end{document}